\documentclass[sn-mathphys-num]{sn-jnl}

\usepackage{graphicx}%
\usepackage{multirow}%
\usepackage{amsmath,amssymb,amsfonts}%
\usepackage{amsthm}%
\usepackage{mathrsfs}%
\usepackage[title]{appendix}%
\usepackage{xcolor}%
\usepackage{textcomp}%
\usepackage{manyfoot}%
\usepackage{booktabs}%
\usepackage{algorithm}%
\usepackage{algorithmicx}%
\usepackage{algpseudocode}%
\usepackage{listings}%

\newcommand{\bkp}{\boldsymbol{\kappa}}

\begin{document}

\title[Hubbard model supplemented with magnetic field]{Density of states of the Hubbard model\\ supplemented with the quantizing magnetic field}%

\author{\fnm{Alexei} \sur{Sherman}}\email{alekseis@ut.ee}

\affil{\orgdiv{Institute of Physics}, \orgname{University of Tartu}, \orgaddress{\street{W. Ostwaldi Str 1}, \city{Tartu}, \postcode{50411}, \country{Estonia}}}

\abstract{Using the strong coupling diagram technique, we calculate the zero-temperature density of states $\rho$ of electrons on a square lattice immersed in a perpendicular uniform magnetic field. The electrons are described by Hubbard Hamiltonian. For moderate doping, Landau subbands are observed for small Hubbard repulsions $U$ only. For larger $U$, the subbands are blurred. Instead, small peaks varying with the field induction $B$ arise by opening the Mott gap in its vicinity. The related variation of $\rho$ with $1/B$ may be connected with the low-frequency quantum oscillations in lightly doped cuprates. For all considered repulsions, $\rho$ has gaps near transfer frequencies of the Hubbard atom, $-\mu$ and $U-\mu$, with $\mu$ the chemical potential. In the heavily underdoped case $\mu<0$, Landau subbands are grouped into the lower and upper Hubbard subbands for moderate and large repulsions. The intensity of the upper Hubbard subband decreases with approaching the Fermi level to the lower edge of the spectrum and finally vanishes.}


\maketitle

\section{Introduction}
External magnetic fields are known to affect the transport properties of crystals. Experimental manifestations of this influence were observed in metals by Shubnikov, de Haas, and van Alphen as oscillations of resistivity and magnetization caused by varying magnetic fields \cite{Shoenberg}. The oscillations are connected with the magnetic splitting of the electron spectrum into Landau levels \cite{Landau} transformed into subbands in crystals \cite{Brown,Langbein}. Closely related to these effects are plateaux in the Hall resistivity of the two-dimensional (2D) electron gas \cite{Klitzing,Thouless}, which are manifestations of the integer quantum Hall effect. The Coulomb interaction between electrons results in further splitting of the Landau subbands. This event was termed the fractional quantum Hall effect \cite{Tsui,Laughlin}. The Coulomb repulsion between electrons is also responsible for the anomalously low frequency of the Shubnikov-de Haas oscillations in lightly doped cuprate perovskites \cite{Sebastian}. One of the possible explanations of these anomalous frequencies relates them to small Fermi-surface pockets located in the nodal region of the Brillouin zone \cite{Sebastian}. Such small pockets arise with the opening of the Mott gap in its vicinity \cite{Sherman15b,Sherman22}.

Presumably, some of the above effects may be understood in the framework of the Hubbard model supplemented with a uniform magnetic field. Calculations using this model were already performed for explaining low-frequency quantum oscillations in lightly doped cuprates in Refs.~\cite{Sherman15b,Sherman22} and for reproducing Shubnikov-de Haas conductivity oscillations in crystals with large Fermi surfaces \cite{Markov,Vucicevic}. The above effects reveal themselves in the model density of states (DOS) $\rho(\omega)$. This work aims to study the DOS in different parameter regions. Since the effects are usually observed at extremely low temperatures $T$, we consider the case $T=0$. We use the strong coupling diagram technique (SCDT) \cite{Vladimir,Metzner,Pairault,Sherman18}. In this approach, Green's functions are calculated by applying series expansions in kinetic energy powers. The method was developed for the case $U\gg t$, where $U$ and $t$ are the repulsion between electrons residing on the same lattice site and intersite hopping constant, respectively. Self-consistent calculations \cite{Sherman18} show that SCDT gives qualitatively correct results also for $U\approx t$. A numerical continuation of obtained results to real frequencies will blur fine structures in DOS. To perform this procedure analytically, we limit the irreducible part to two lowest-order terms of the SCDT expansion.

We find that at half-filling, $\mu=U/2$, and for moderate doping, Landau subbands are well resolved only for small Hubbard repulsions. Here $\mu$ is the chemical potential. For larger $U$, subbands are blurred and merge together. Instead, small peaks appear near the Mott gap with its opening. Their location depends on the field induction $B$. We relate the peaks to states satisfying the Onsager quantization condition \cite{Shoenberg}. When the Fermi level (FL) falls into the region of the peaks, their dependence on the magnetic field causes oscillations in $\rho(\omega=0)$. The frequency of these oscillations will be anomalously low due to small Fermi-surface pockets for such electron concentrations \cite{Sherman15b,Sherman22}. Hence, the peaks can be related to the low-frequency quantum oscillations observed in lightly doped cuprates \cite{Sebastian}. For all doping and Hubbard repulsions, the zero-temperature DOSs have gaps at the Hubbard atom transfer frequencies $\omega=-\mu$ and $U-\mu$. For $B=0$, finite-temperature DOSs have dips at these frequencies \cite{Grober}. Both zero-temperature gaps and finite-temperature dips stem from multiple reabsorptions of electrons by Hubbard atoms \cite{Sherman18}. The gap widths grow with doping. The DOS is cardinally changed in the heavily underdoped case when $\mu<0$. Here, Landau subbands are well resolved not only for $U\sim t$, but for moderate and large repulsions as well. In the later cases, lower and upper Hubbard subbands are composed of an equal number of Landau subbands. The intensity of the upper Hubbard subband decreases with approaching the FL to the lower spectral edge. This spectral modification follows from the fact that the local interaction between electrons becomes increasingly rare with the reduction of their concentration.

The article is organized as follows: The model Hamiltonian and the main formulas are given in the next section. The results of calculations and their discussion are brought up in Sect. 3. The last section is devoted to concluding remarks.

\section{Model and main formulas}
In the presence of the uniform magnetic field ${\bf B}$ perpendicular to a square crystal lattice, electrons with the on-site Coulomb interaction are described by the Hamiltonian
\begin{eqnarray}\label{Peierls}
H&=&-t\sum_{{\bf la}\sigma}\exp\left({\rm i}\frac{e}{\hbar}\int_{\bf l-a}^{\bf l}{\bf A}({\bf r}){\rm d}{\bf r}\right)a^\dagger_{{\bf l-a},\sigma}a_{{\bf l}\sigma}+\frac{1}{2}g\mu_BB \sum_{{\bf l}\sigma}\sigma n_{{\bf l}\sigma}\nonumber\\
&&+U\sum_{\bf l}n_{{\bf l}\uparrow}n_{{\bf l},\downarrow},
\end{eqnarray}
where the 2D vector ${\bf l}$ labels sites of the lattice, ${\bf a}$ are four vectors connecting nearest neighbor sites, $\sigma=\uparrow,\downarrow$ is the spin projection, $e$ and $\hbar$ are the modulus of the electron charge and Planck constant, respectively, ${\bf A}({\bf r})$ is the vector potential, $a^\dagger_{{\bf l}\sigma}$ and $a_{{\bf l}\sigma}$ are electron creation and annihilation operators, $g$ and $\mu_B$ are the $g$-factor and Bohr magneton, respectively, and $n_{{\bf l}\sigma}=a^\dagger_{{\bf l}\sigma}a_{{\bf l}\sigma}$ is the electron number operator.

In Eq.~(\ref{Peierls}), the influence of the magnetic field on the kinetic term is described in the Peierls approximation \cite{Peierls,Wannier}. It is valid until the magnetic length $l_B=\sqrt{\hbar/(eB)}$ is larger than the size of the Wannier function $l_W$. Below, we shall use strong fields to decrease computation time and promote the convergence of the used iteration procedure. For such fields, $l_B$ is only slightly larger than $a=|{\bf a}|$. Nevertheless, we suppose that the condition $l_W<l_B$ is fulfilled. We also assume that the Zeeman splitting described by the next to the last term is much smaller than the energy parameters of the Hamiltonian and widths of spectral features. Therefore, this term will be omitted below.

Using the Landau gauge \cite{Landau} ${\bf A}({\bf l})=-Bl_y{\bf x}$, the exponent in the kinetic term in (\ref{Peierls}) can be written as $-{\rm i}eBl_ya_x/\hbar$. We suppose that the lattice is located in the $x$-$y$ plane, ${\bf B}$ is directed along the $z$ axis, ${\bf x}$ is the unit vector along the $x$ axis, $l_y$ and $a_x$ are $y$ and $x$ components of the respective vectors. We restrict ourselves to fields satisfying the condition
\begin{equation}\label{condition}
\frac{e}{\hbar}Ba^2=2\pi\frac{\nu'}{\nu},
\end{equation}
where $\nu'$ and $\nu$ are coprime integers. With these notations and approximations, the Hamiltonian reads
\begin{equation}\label{Hamiltonian}
H=-t\sum_{{\bf la}\sigma}\exp\left(-{\rm i}\frac{\nu'a_x}{a}\bkp{\bf l}\right)a^\dagger_{{\bf l-a},\sigma}a_{{\bf l}\sigma}+U\sum_{\bf l}n_{{\bf l}\uparrow}n_{{\bf l},\downarrow},
\end{equation}
where $\bkp=2\pi/(\nu a){\bf y}$. As seen from Eq.~(\ref{Hamiltonian}), the Hamiltonian is invariant with respect to translations by the lattice period along the $x$ axis and by $\nu$ lattice periods along the $y$ axis. Below, we use a lattice containing $N_x$ sites along the $x$ axis and $\nu N_y$ sites along the $y$ axis with the periodic boundary conditions. In such a lattice, the following Fourier transformation over the spacial variables can be defined:
\begin{equation}\label{Fourier}
a^\dagger_{{\bf k}m\sigma}=\frac{1}{\sqrt{N}}\sum_{\bf l}{\rm e}^{{\rm i}({\bf k}+m\bkp){\bf l}} a^\dagger_{{\bf l}\sigma},
\end{equation}
where the wave vector ${\bf k}=(2\pi n_x/(N_xa),2\pi n_y/(\nu N_ya))$, $n_x=0,1,\ldots N_x-1$, $n_y=0,1,\ldots N_y-1$ is defined in the reduced, magnetic Brillouin zone, $m=0,1,\ldots\nu-1$, and $N=\nu N_xN_y$. $m$ is a cyclic variable with the period $\nu$.

Below, we calculate the DOS
\begin{equation}\label{DOS}
\rho(\omega)=-\frac{1}{\pi N}\sum_{{\bf k}m}{\rm Im}\,G_{mm}({\bf k},\omega),
\end{equation}
where $G_{mm}({\bf k},\omega)$ is the Fourier transform of Green's function $G({\bf l'}\tau',{\bf l}\tau)=\langle{\cal T}\bar{a}_{{\bf l'}\sigma}(\tau')a_{{\bf l}\sigma}(\tau)\rangle$ after the analytic continuation to the real frequency axis. Here angle brackets denote the statistical averaging with the operator ${\cal H}=H-\mu\sum_{{\bf l}\sigma}n_{{\bf l}\sigma}$, which defines also time dependencies of operators in $G({\bf l'}\tau',{\bf l}\tau)$, ${\cal T}$ is the chronological operator. To calculate this Green's function, we use the SCDT series expansion over the powers of the kinetic energy in the Hamiltonian (\ref{Hamiltonian}). Terms of the expansion are products of the hopping constants with the Peierls exponential $t_{\bf ll'}=-t\sum_{\bf a}\exp(-{\rm i}\nu'a_x\bkp{\bf l}/a)\delta_{\bf l',l-a}$ and on-site cumulants of electron creation and annihilation operators. The terms of the expansion can be visualized by depicting $t_{\bf ll'}$ as directed lines and cumulants as circles with the number of outgoing and incoming lines equal to the number of creation and annihilation operators in them. The linked-cluster theorem is valid, and partial summations are allowed in SCDT. A two-leg diagram is irreducible if it cannot be divided into two disconnected parts by cutting a hopping line $t_{\bf ll'}$. Denoting the sum of all such diagrams by ${\bf K}$, the Fourier transform of the electron Green's function is written as
\begin{equation}\label{Larkin}
{\bf G}({\bf k},\omega)=\left\{\left[{\bf K}({\bf k},\omega)\right]^{-1}-{\bf t_k}\right\}^{-1} =\left[{\bf 1}-{\bf K}({\bf k},\omega){\bf t_k}\right]^{-1}{\bf K}({\bf k},\omega),
\end{equation}
where ${\bf t_k}$ is the Fourier transform of $t_{\bf ll'}$,
\begin{equation}\label{hopping}
t_{mm'}({\bf k})=-t\left[{\rm e}^{-{\rm i}k_xa}\delta_{m',m+\nu'}+{\rm e}^{{\rm i}k_xa}\delta_{m',m-\nu'}+2\cos\left(k_ya+\frac{2\pi m}{\nu}\right)\delta_{m,m'}\right].
\end{equation}
Notice that after Fourier transformations, $G$, $K$, and $t$ depend on two variables $m$ and $m'$, which are considered as matrix indices in Eq.~(\ref{Larkin}).

\begin{figure}[t]
\centerline{\resizebox{0.8\columnwidth}{!}{\includegraphics{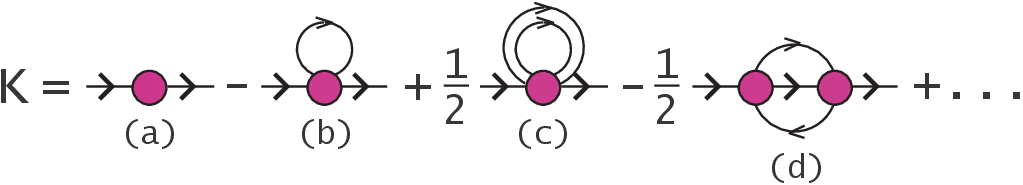}}}
\caption{Diagrams of the first three orders in ${\bf K}({\bf k},\omega)$.} \label{Fig1}
\end{figure}
Diagrams of the first three orders in ${\bf K}({\bf k},\omega)$ are shown in Fig.~\ref{Fig1}. Thanks to the possibility of carrying out partial summations, internal lines ${\bf t_k}$ in these diagrams can be substituted by the renormalized hopping
\begin{equation}\label{theta}
\boldsymbol{\theta}({\bf k},\omega)={\bf t_k}+{\bf t_k}{\bf G}({\bf k},\omega){\bf t_k}.
\end{equation}

In the below calculations, the irreducible part ${\bf K}({\bf k},\omega)$ is approximated by the sum of diagrams (a) and (b) in Fig.~\ref{Fig1}. As seen below, this approximation correctly describes the main peculiarities of DOS shapes in the absence of the magnetic field. It gives the critical repulsion for the Mott gap opening, $U_c\approx6t$, agreeing with results from an infinite series of SCDT ladder diagrams \cite{Sherman18}. The main difference between DOSs in these two approximations, neglected in the present work, is the narrow Slater gap at half-filling near $\omega=0$ for moderate $U$ \cite{Sherman23}. The gap is a consequence of the spin antiferromagnetic ordering. In the used approximation, ${\bf K}$ is local. Therefore, it does not describe any spin ordering, and the Slater gap does not appear in DOSs. Nevertheless, we use this approximation because it allows us to perform the continuation to real frequencies analytically without recourse to the maximum entropy method. The irreducible part reads
\begin{eqnarray}\label{K1}
K_{m'm}(j)&=&C^{(1)}(j)\delta_{m',m}-\frac{T}{N}\sum_{j'\sigma'}\sum_{{\bf k}m''} C^{(2)}(j\sigma,j\sigma,j'\sigma',j'\sigma')\theta_{m'',m-m'+m''}({\bf k},j') \nonumber\\
&=& C^{(1)}(j)\delta_{m',m}-T\sum_{j'\sigma'}C^{(2)}(j\sigma,j\sigma,j'\sigma',j'\sigma') \varphi_{m'-m}(j'),
\end{eqnarray}
where $C^{(1)}$ and $C^{(2)}$ are cumulants of the first and second orders, respectively, $j$ is an integer defining the fermion Matsubara frequency $\omega_j=(2j-1)\pi T$,
\begin{eqnarray}\label{varphi}
\varphi_m(j)&=&\frac{1}{N}\sum_{{\bf k}m_1}\sum_{m_2m_3}t_{m_1m_2}({\bf k})G_{m_2m_3}({\bf k},j)t_{m_3,m_1-m}({\bf k})\nonumber\\
&=&\frac{1}{N}\sum_{{\bf k}m_1m_2}\left[{\bf 1}-{\bf t}({\bf k}){\bf K}(j)\right]^{-1}\!\rule[-1.5mm]{0.1mm}{4.5mm}_{\,m_1,m_2}t_{m_2,m_1-m}({\bf k}).
\end{eqnarray}
In moving from the first line to the second in (\ref{K1}), we used Eq.~(\ref{theta}) and took into account that $t_{\bf ll}=0$. Notice that in the used approximation, $K_{m'm}(j)$ depends only on the difference $m'-m$.

For the analytic continuation to the real frequency axis, we need to find a relation between the quantity $\varphi_m(j)$ on the upper half of the imaginary axis and its values infinitesimally above the real axis. From the spectral representation, one can see that
\begin{equation*}
G_{m'm}({\bf k},j)=\frac{1}{2\pi{\rm i}}\int_{-\infty}^{\infty}\frac{G_{m'm}({\bf k},\omega+{\rm i}\eta)-G^*_{mm'}({\bf k},\omega+{\rm i}\eta)}{\omega-{\rm i}\omega_j}\,{\rm d}\omega,\quad \eta\rightarrow+0.
\end{equation*}
Taking into account that $t_{mm'}({\bf k})$ is a Hermitian matrix, $t^*_{m'm}({\bf k})=t_{mm'}({\bf k})$, we find from this equation and (\ref{varphi})
\begin{equation}\label{imagreal}
\varphi_{m}(j)=\frac{1}{2\pi{\rm i}}\int_{-\infty}^{\infty}\frac{\varphi_{m}(\omega+{\rm i}\eta)-\varphi^*_{-m}(\omega+{\rm i}\eta)}{\omega-{\rm i}\omega_j}\,{\rm d}\omega.
\end{equation}

Expressions for cumulants appearing in (\ref{K1}) read \cite{Vladimir,Metzner,Pairault,Sherman18}
\begin{eqnarray}
&&C^{(1)}(j)=Z^{-1}\left[\left({\rm e}^{-\beta E_0}+{\rm e}^{-\beta E_1}\right)g_1({\rm i}\omega_j)+ \left({\rm e}^{-\beta E_1}+{\rm e}^{-\beta E_2}\right)g_2({\rm i}\omega_j)\right], \nonumber\\
&&\sum_{\sigma'}C^{(2)}(j\sigma,j\sigma,j'\sigma',j'\sigma')=\Big[Z^{-1}{\rm e}^{-\beta E_1}\left(1-2\delta_{j,j'}\right)\nonumber\\
&&\quad+Z^{-2}\left({\rm e}^{-\beta(E_0+E_2)}-{\rm e}^{-2\beta E_1}\right)\left(2-\delta_{j,j'}\right)\Big]\beta F({\rm i}\omega_j)F({\rm i}\omega_{j'})\nonumber\\
&&\quad-Z^{-1}{\rm e}^{-\beta E_0}Ug_1({\rm i}\omega_j)g_1({\rm i}\omega_{j'})g_3({\rm i}\omega_j+{\rm i}\omega_{j'})\left[g_1({\rm i}\omega_j)+g_1({\rm i}\omega_{j'})\right] \label{cumulants}\\
&&\quad-Z^{-1}{\rm e}^{-\beta E_2}Ug_2({\rm i}\omega_j)g_2({\rm i}\omega_{j'})g_3({\rm i}\omega_j+{\rm i}\omega_{j'})\left[g_2({\rm i}\omega_j)+g_2({\rm i}\omega_{j'})\right]\nonumber\\
&&\quad+Z^{-1}{\rm e}^{-\beta E_1}\big\{F({\rm i}\omega_{j'})\left[g_1^2({\rm i}\omega_j)-F({\rm i}\omega_j)g_2({\rm i}\omega_{j'})\right]\nonumber\\
&&\quad+F({\rm i}\omega_j)\left[g_1^2({\rm i}\omega_{j'})
-F({\rm i}\omega_{j'})g_2({\rm i}\omega_j)\right]\big\},\nonumber
\end{eqnarray}
where $E_0=0$, $E_1=-\mu$, and $E_2=U-2\mu$ are eigenenergies of the Hubbard atom Hamiltonian, $H_{\bf l}=Un_{{\bf l}\uparrow}n_{{\bf l}\downarrow}-\mu\sum_\sigma n_{{\bf l}\sigma}$, corresponding to the empty, singly, and doubly occupied states, $\beta=1/T$, $Z=\exp(-\beta E_0)+2\exp(-\beta E_1)+\exp(-\beta E_2)$,
\begin{eqnarray*}
&&g_1({\rm i}\omega_j)=({\rm i}\omega_j+\mu)^{-1},\quad g_2({\rm i}\omega_j)=({\rm i}\omega_j+\mu-U)^{-1},\\
&&g_3({\rm i}\omega_j+{\rm i}\omega_{j'})=({\rm i}\omega_j+{\rm i}\omega_{j'}+2\mu-U)^{-1},\quad F({\rm i}\omega_j)=g_1({\rm i}\omega_j)-g_2({\rm i}\omega_j).
\end{eqnarray*}
Substituting cumulants (\ref{cumulants}) and Eq.~(\ref{imagreal}) into (\ref{K1}), performing the summation over $j'$, replacing ${\rm i}\omega_j$ with $\omega+{\rm i}\eta$, and setting $T=0$, we find
\begin{eqnarray}\label{K2}
K_m(\omega)&=&H(-\mu)\bigg\{g_1(\omega+{\rm i}\eta)\delta_{m,0}-\frac{U}{2\pi{\rm i}}\int_{-\infty}^{0} \left[\varphi_m(\omega')-\varphi^*_{-m}(\omega')\right] \nonumber\\
&&\quad\times g_1(\omega)g_1(\omega')g_3(\omega+\omega')\left[g_1(\omega)+g_1(\omega')\right] \bigg\}{\rm d}\omega' \nonumber\\
&+&H(\mu)H(U-\mu)\bigg\{\frac{1}{2}\left[g_1(\omega)+g_2(\omega)\right]\delta_{m,0} +\frac{3}{4}F^2(\omega)\varphi_m(\omega)\nonumber\\
&&\quad-\frac{s^{(1)}_m}{2}\left[g_1^2(\omega)- F(\omega)g_2(\omega)\right]-\frac{s^{(2)}_m}{2}F(\omega)\bigg\}\nonumber\\
&+&H(\mu-U)\bigg\{g_2(\omega+{\rm i}\eta)\delta_{m,0}+\frac{U}{2\pi{\rm i}}\int_{0}^{\infty}\left[\varphi_m(\omega')-\varphi^*_{-m}(\omega')\right] \nonumber\\
&&\quad\times g_2(\omega)g_2(\omega')g_3(\omega+\omega')\left[g_2(\omega)+g_2(\omega')\right] \bigg\}{\rm d}\omega',
\end{eqnarray}
where $H(\omega)$ is the Heaviside step function,
\begin{eqnarray}
s^{(1)}_m&=&\frac{1}{2\pi{\rm i}}\int_{-\infty}^{\infty}\left[\varphi_m(\omega)-\varphi^*_{-m}(\omega)\right]\left[H(\omega) g_1(\omega)+H(-\omega)g_2(\omega)\right]{\rm d}\omega,\nonumber\\
s^{(2)}_m&=&\frac{1}{2\pi U{\rm i}}\int_{-\infty}^{\infty}\left[\varphi_m(\omega)-\varphi^*_{-m}(\omega)\right]\big[H(\omega) (\omega+\mu+U)g^2_1(\omega) \label{s_i}\\
&&+H(-\omega)(\omega+\mu-2U)g^2_2(\omega)\big]{\rm d}\omega.\nonumber
\end{eqnarray}

Notice that the irreducible part $K_m(\omega)$ is different in the region of half-filling and moderate doping $0<\mu<U$ and for the heavily underdoped $\mu<0$ and overdoped $\mu>U$ cases. It is a consequence of the fact that in these three regions, the ground state of the Hubbard atom Hamiltonian $H_{\bf l}$ is different. With growing $\mu$, it is changed from the empty to two singly occupied and finally to the doubly occupied state. Various ground states lead to differences in the cumulants and irreducible parts in the mentioned $\mu$ regions. This discontinuous change is a consequence of zero temperature and electron correlations. It is not a property inherent solely in SCDT. For example, similar abrupt changes in Green's function with $\mu$ \cite{Sherman22} occur in the cluster perturbation theory (CPA) \cite{Senechal02}. In this approach, the crystal Green's function is calculated from the cluster ones. The latter have many ground states corresponding to different $\mu$. We are interested in three regions of $\mu$ and two abrupt changes in Green's functions in these regions. We are talking about the region with the electron concentration $x=1$ and two of its neighbors with the number of electrons one less or more than the number of cluster sites $N_c$. These three regions are analogous to the above $\mu$ regions of the Hubbard atom. The difference between the three cluster regions is retained even for $N_c\rightarrow\infty$ when the CPA becomes exact. It occurs because the regions are qualitatively dissimilar -- one corresponds to an insulator and two other metals. We suppose that the difference persists even for $U<U_c$ until $U\gtrsim t$. Accordingly, the abrupt change of Green's function is also maintained at $T=0$ in this case.

The Hamiltonian (\ref{Hamiltonian}) has the particle-hole symmetry. To see this, we change the electron operators as follows:
\begin{equation*}
a_{{\bf l}\sigma}={\rm e}^{{\rm i}{\bf Ql}}\tilde{a}^\dagger_{{\bf l}\sigma},\quad a^\dagger_{{\bf l}\sigma}={\rm e}^{-{\rm i}{\bf Ql}}\tilde{a}_{{\bf l}\sigma},\quad {\bf Q}=\left(\frac{\pi}{a},\frac{\pi}{a}\right),
\end{equation*}
reverse the field direction, and take into account that $a_xa_y=0$. Therefore, in the following consideration, we only consider chemical potentials $\mu\leq U/2$.

Equations (\ref{hopping}), (\ref{varphi}), (\ref{K2}), and (\ref{s_i}) form a closed system, which can be solved by iteration for given values of $U/t$, $\mu/t$, $\nu$, and $\nu'$. In this procedure, as the starting values of $K_m(\omega)$, we used the first terms in braces in Eq.~(\ref{K2}). To achieve the proper analytic properties of the retarded Green's function, a small imaginary constant was added to the frequency in the functions $g_1(\omega)$ and $g_2(\omega)$ in the starting $K_m(\omega)$, $\omega\rightarrow\omega+{\rm i}\eta$, $\eta\sim0.01t$. For the case $0<\mu\leq U/2$, no other artificial broadening was used in the following iteration steps. In the case $\mu<0$, we had to add such a broadening in the first term in braces for all iteration steps to achieve convergence. After the iteration convergence, the obtained irreducible part was applied for calculating Green's function (\ref{Larkin}) and the DOS (\ref{DOS}).

\section{Results and discussion}
\begin{figure}[t]
\centerline{\resizebox{0.8\columnwidth}{!}{\includegraphics{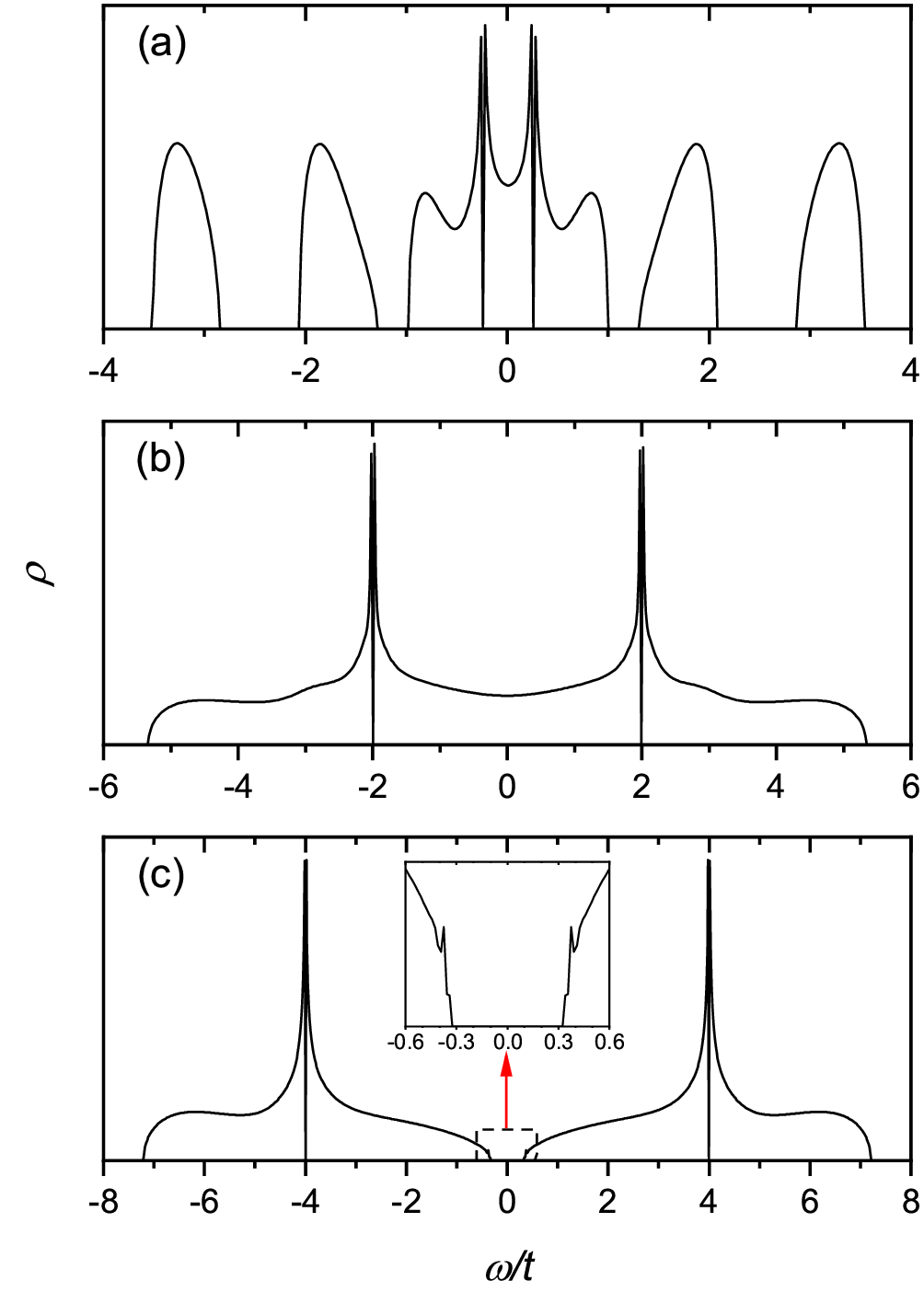}}}
\caption{Densities of states for half-filling, $U=0.5t$, $\nu=7$ (a), $U=4t$, $\nu=5$ (b), and $U=8t$, $\nu=3$ (c). For all panels $\nu'=1$. The inset in panel (c) shows the vicinity of the Mott gap on an increased scale.} \label{Fig2}
\end{figure}
Densities of states calculated at half-filling, $x=2\int_{-\infty}^{0}\rho(\omega){\rm d}\omega=1$, for several values of the Hubbard repulsion $U$ and strengths of the magnetic field are shown in Fig.~\ref{Fig2}. For $U=0.5t$, all seven (for $\nu=7$) Landau subbands are well resolved, though three central ones merge. However, at $U=4t$, gaps between Landau subbands disappear, and only weak remnant DOS variations near $\omega=\pm3t$ remind of them.

\begin{figure}[t]
\centerline{\resizebox{0.8\columnwidth}{!}{\includegraphics{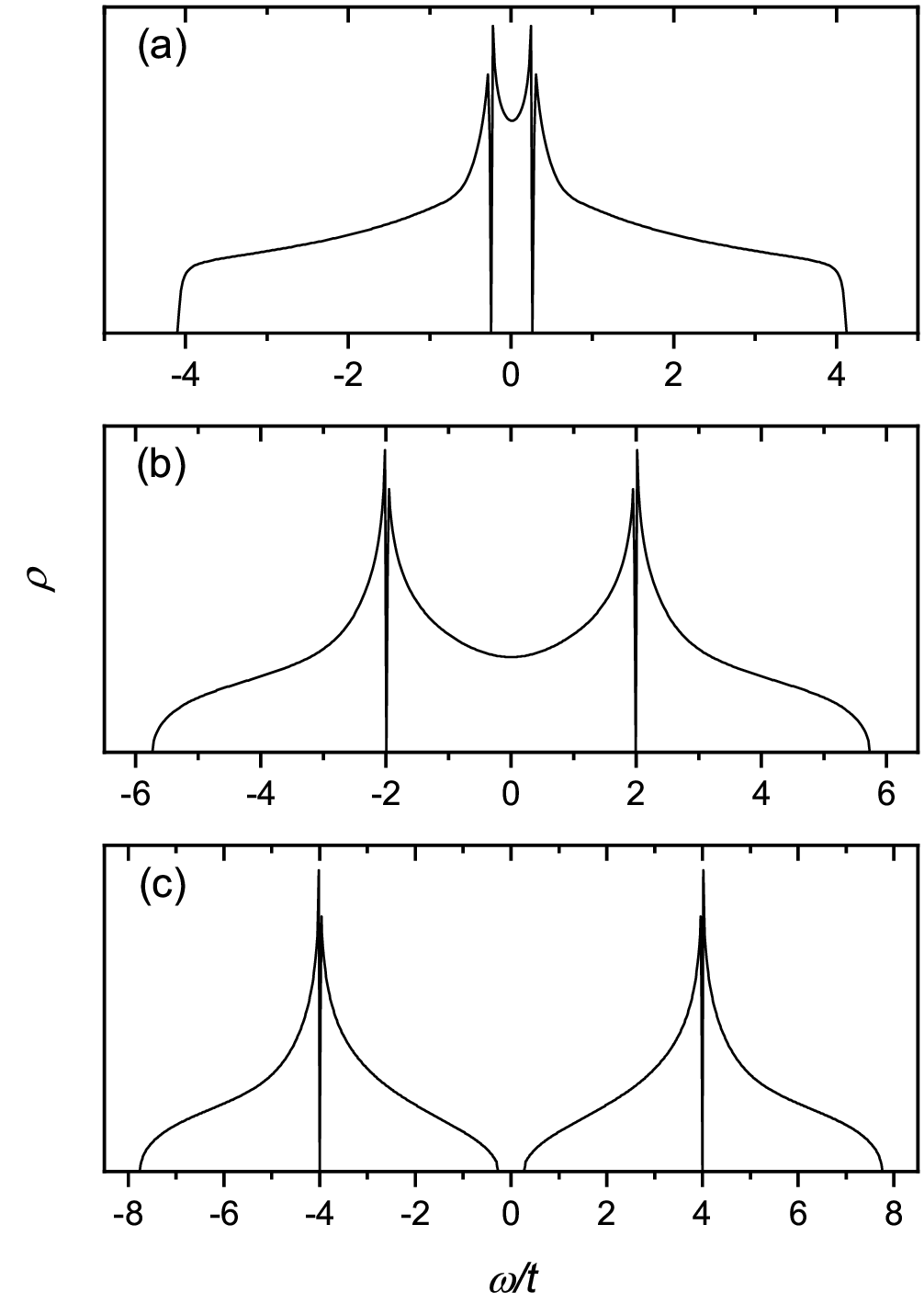}}}
\caption{Densities of states for half-filling, $B=0$, $U=0.5t$ (a), $U=4t$ (b), and $U=8t$ (c).} \label{Fig3}
\end{figure}

For $U=4t$ and $8t$, the DOS shapes are close to those calculated with the same approximation for $B=0$. For comparison, these later DOSs are shown in Fig.~\ref{Fig3}.
Equations for calculating these spectra are similar to those in the previous section with the following differences: Matrix equations (\ref{Larkin}) and (\ref{theta}) transform to the scalar form with ${\bf k}$ the wave vector in the usual Brillouin zone. Only the $m=0$ components are retained in Eqs.~(\ref{DOS}), (\ref{varphi}), (\ref{K2}), (\ref{s_i}), and the hopping function (\ref{hopping}) transforms to $t_{\bf k}=-2t[\cos(k_xa)+\cos(k_ya)]$. As follows from the comparison of zero-temperature spectra in Fig.~\ref{Fig3} with finite-temperature results of Ref.~\cite{Sherman18} obtained with an infinite series of SCDT diagrams, the used approximation qualitatively correctly reproduces the main features of DOSs in the square-lattice Hubbard model. The main differences between the two approximations are located near frequencies $\omega=-\mu$, $U-\mu$, and $0$. In the finite-temperature calculations, the DOS is suppressed near the transfer frequencies of the Hubbard atom due to the electron reabsorption, leading to the spectrum's four-band structure. It is observed also in Monte Carlo simulations \cite{Grober}. We obtained narrow gaps at these frequencies instead of the intensity suppressions in the zero-temperature calculations. As the above formulas show, the irreducible part does not depend on the wave vector. As a consequence, the used approximation does not describe any spin ordering neither for itinerant electrons for small $U$ (the Slater mechanism \cite{Slater}) nor for localized electrons at larger repulsions (the Heisenberg mechanism). Therefore, there is no conical Slater gap at $\omega=0$ in Figs.~\ref{Fig3}(a) and (b), which is present in calculations describing the momentum dependence of the irreducible part \cite{Sherman23}. Notice, however, that the used approximation is enough to estimate correctly the critical repulsion of the Mott transition $U_c\approx 6t$, which is in agreement with more exact SCDT calculations \cite{Sherman18,Sherman23}.

\begin{figure}[t]
\centerline{\resizebox{0.8\columnwidth}{!}{\includegraphics{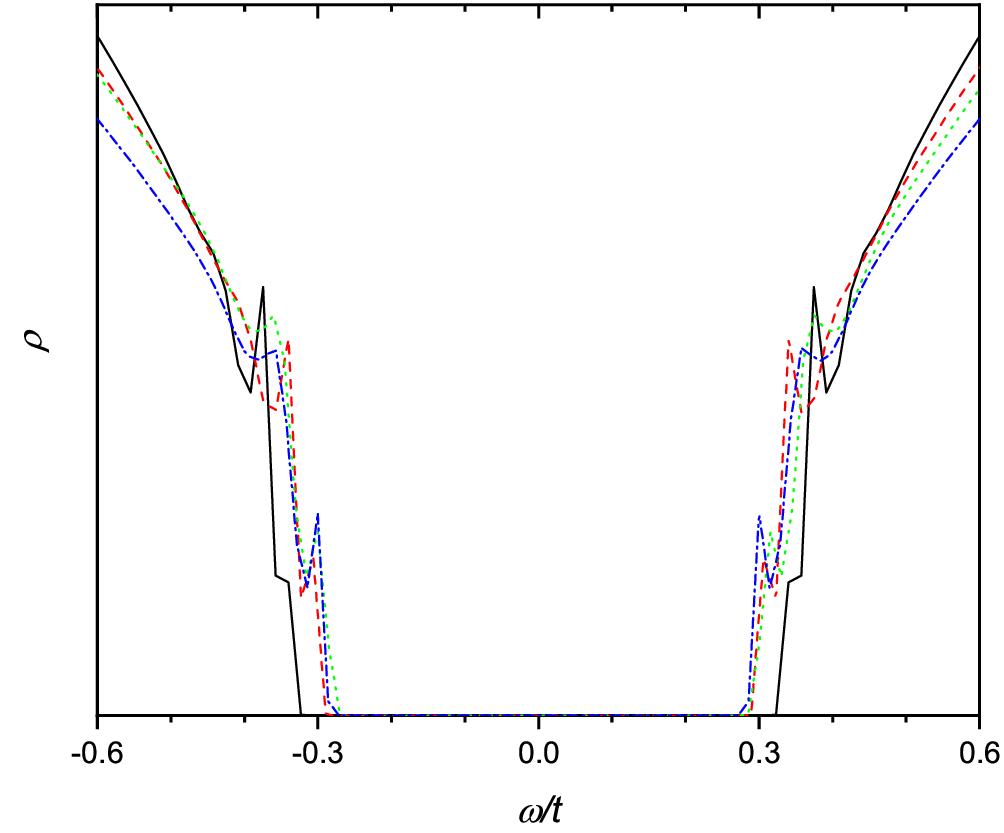}}}
\caption{Densities of states near the Mott gap for half-filling, $U=8t$, and $\nu=3$ (black solid line), $5$ (red dashed line), $9$ (green dotted line), $20$ (blue dash-dotted line). For all curves $\nu'=1$.} \label{Fig4}
\end{figure}
Comparing Figs.~\ref{Fig2}(c) and \ref{Fig3}(c), we see that the magnetic field leads to the appearance of weak maxima near the Mott gap. The location of these peaks depends on the value of $B$, as demonstrated in Fig.~\ref{Fig4}. The Mott gap and peaks are retained at moderate doping. Hence, when the chemical potential is located in the peak region, $\rho(\omega=0)$ will oscillate with $B$. The DOS on FL is the second derivative of the Landau thermodynamic potential $\Omega(T,V,\mu,B)$ over $\mu$ ($V$ is the volume). Therefore, the oscillation of $\rho(0)$ points to the similar variation of $\Omega$ with $B$.
\begin{figure}[t]
\centerline{\resizebox{0.8\columnwidth}{!}{\includegraphics{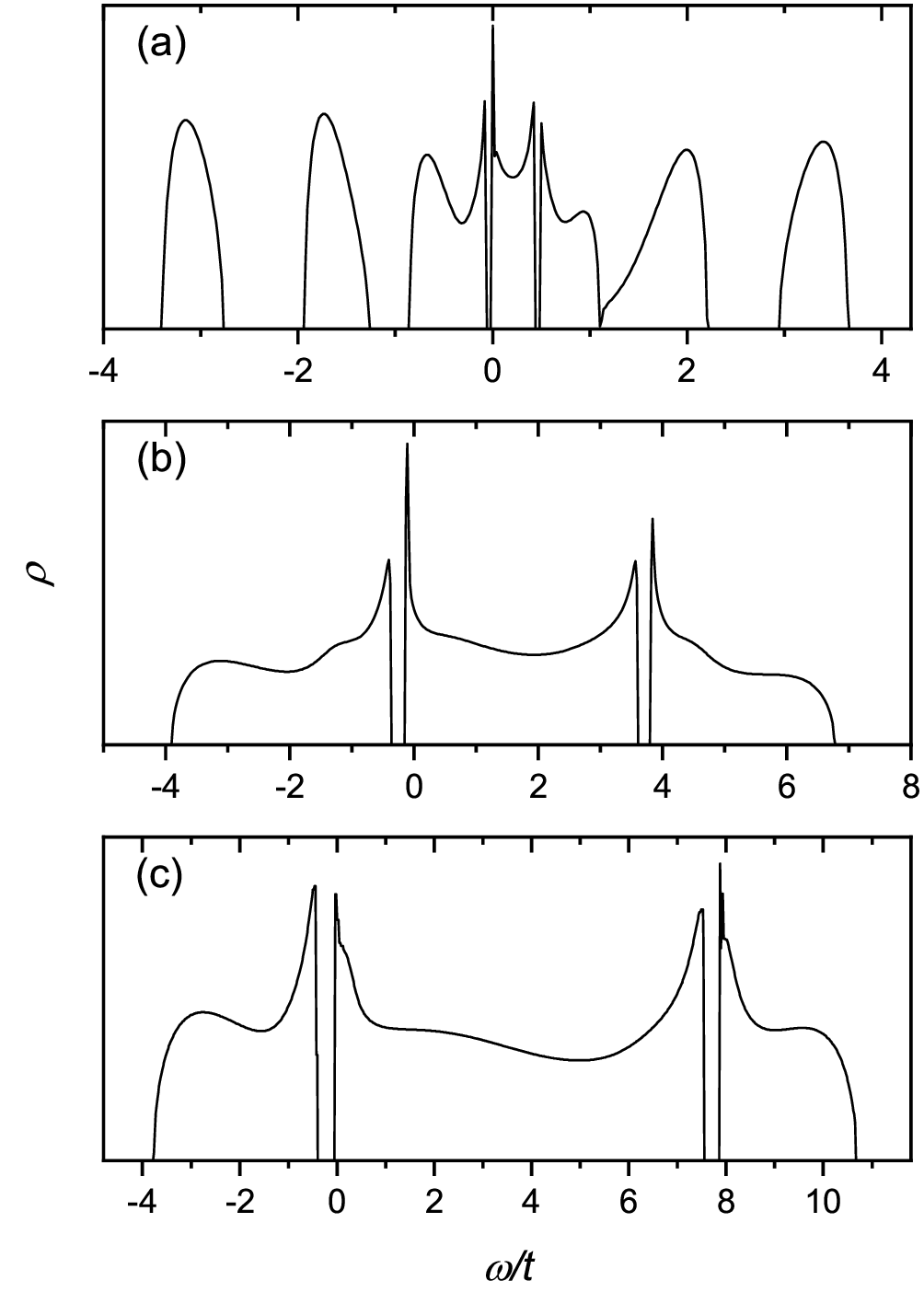}}}
\caption{Densities of states for $U=0.5t$, $\mu=0.05t$, $\nu=7$ ($x=0.95$, a), $U=4t$, $\mu=0.3t$, $\nu=5$ ($x=0.7$, b), and $U=8t$, $\mu=0.3t$, $\nu=3$ ($x=0.53$, c). For all curves $\nu'=1$.} \label{Fig5}
\end{figure}
Since quantum oscillation experiments measure changes of different derivatives of this thermodynamic potential with the magnetic field strength \cite{Shoenberg}, the variation of the peaks with $B$ is the source of these oscillations. We relate the peaks to electron states satisfying the Onsager quantization condition \cite{Shoenberg}. Opening the Mott gap leads to the appearance of small Fermi surface pockets in the vicinity of the gap \cite{Sherman22,Sherman15b}. The consequence of their small area is anomalously low frequencies of quantum oscillations, which are close to those observed in lightly doped cuprates \cite{Sebastian}.
\begin{figure}[t]
\centerline{\resizebox{0.8\columnwidth}{!}{\includegraphics{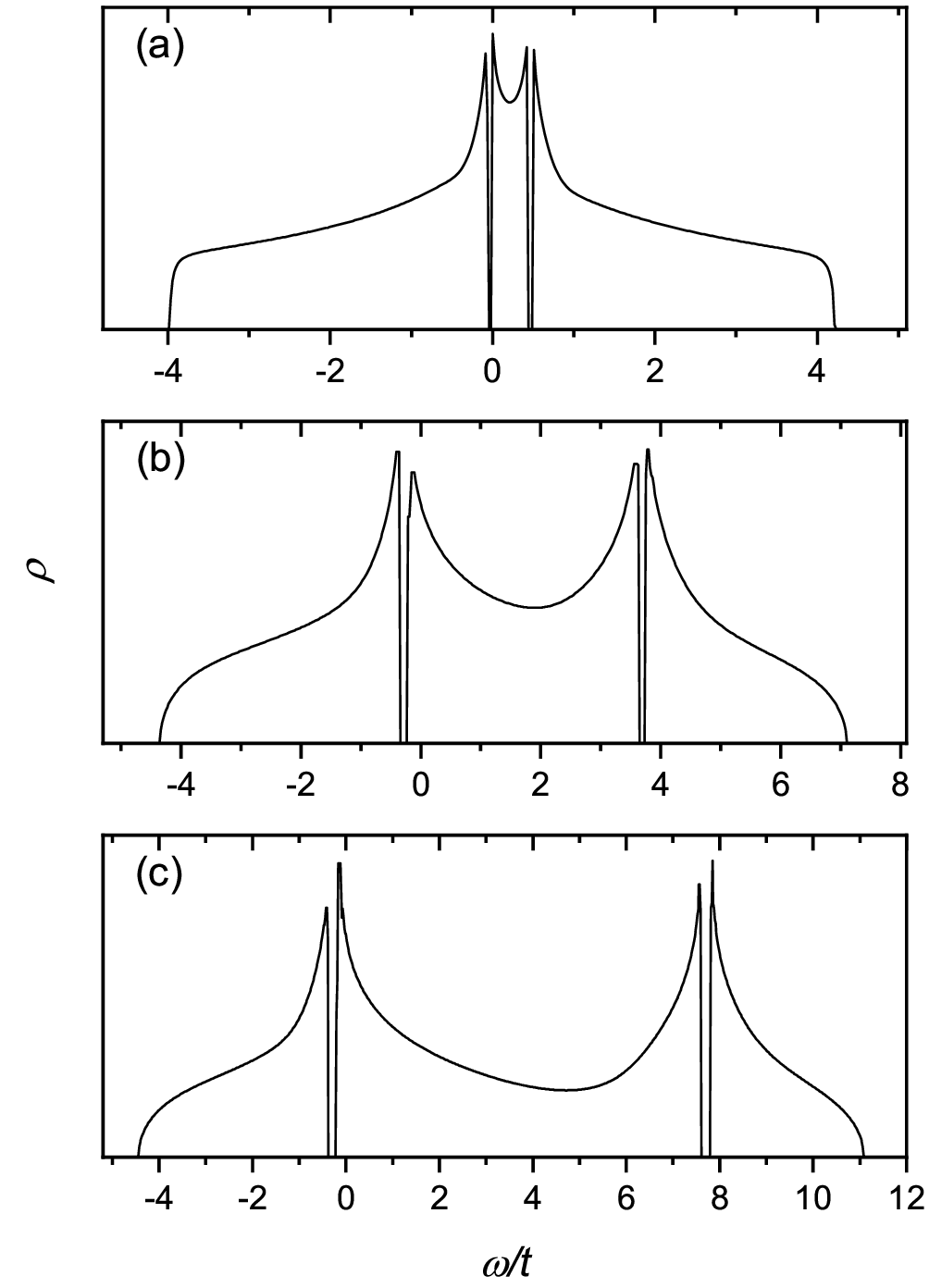}}}
\caption{Densities of states for $B=0$, $U=0.5t$, $\mu=0.05t$ ($x=0.94$, a), $U=4t$, $\mu=0.3t$ ($x=0.75$, b), and $U=8t$, $\mu=0.3t$ ($x=0.57$, c).} \label{Fig6}
\end{figure}

Figure~\ref{Fig5} demonstrates DOSs in the doped case when the chemical potential is close to the lower edge of the range $0<\mu<U$. As seen, doping leads to the significant widening of gaps near $\omega=-\mu$ and $U-\mu$. Besides, for $U=8t$, the Mott gap disappears. As for half-filling, Landau subbands are well resolved for $U=0.5t$ and reveal themselves in remnant DOS variations for larger repulsions. These spectral features are more noticeable from the comparison with the zero-field DOSs in Fig.~\ref{Fig6}. As seen, the reabsorption gaps are wider for $B\neq0$.

\begin{figure}[t]
\centerline{\resizebox{0.8\columnwidth}{!}{\includegraphics{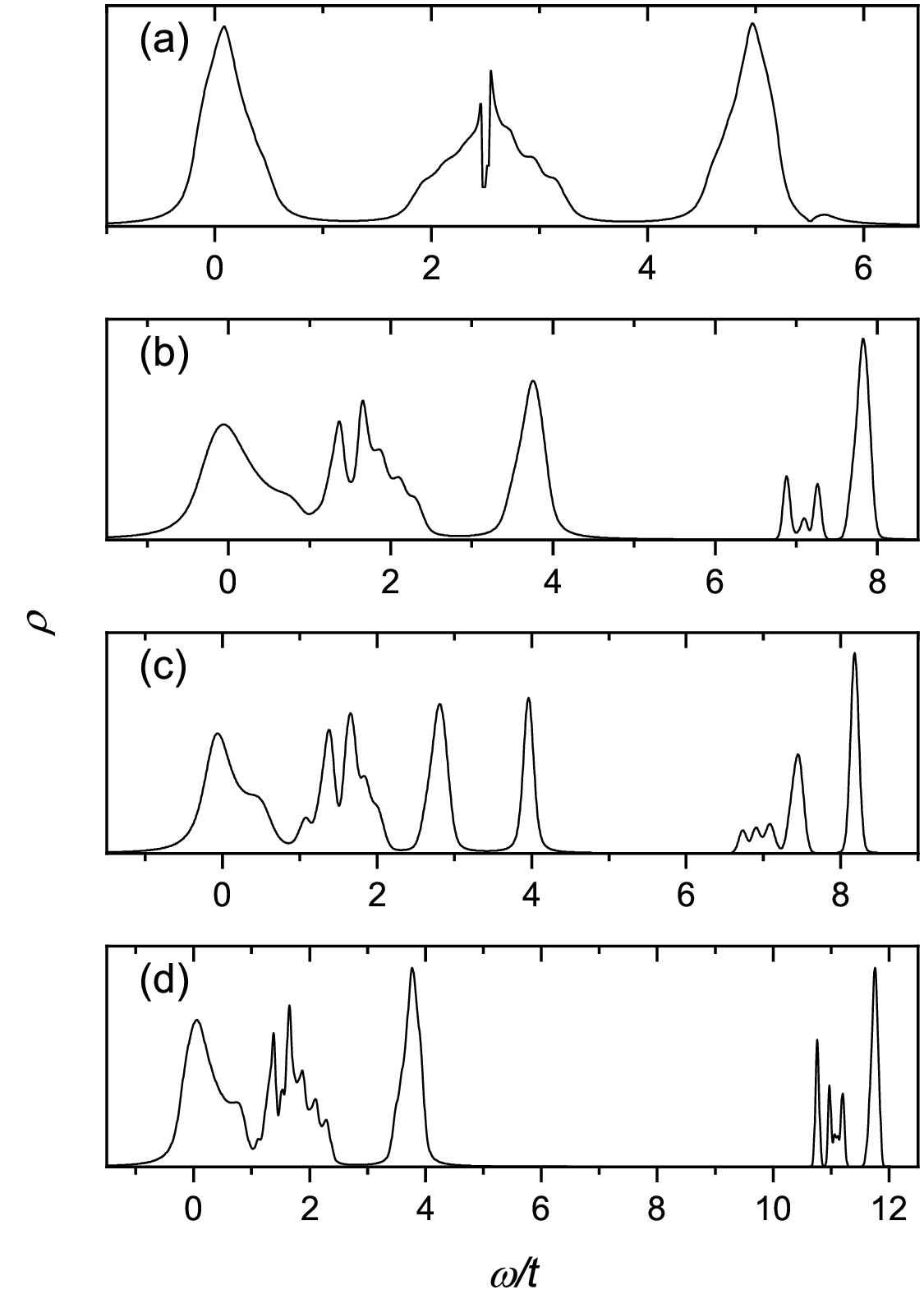}}}
\caption{Densities of states for $U=0.5t$, $\mu=-2.5t$, $\nu=3$ ($x=0.22$, a), $U=4t$, $\mu=-1.5t$, $\nu=3$ ($x=0.3$, b), $U=4t$, $\mu=-1.5t$, $\nu=5$ ($x=0.27$, c), and $U=8t$, $\mu=-1.5t$, $\nu=3$ ($x=0.22$, d). In all cases $\nu'=1$.} \label{Fig7}
\end{figure}
Let us now consider the heavily underdoped case $\mu<0$. As mentioned above, for such chemical potentials, the irreducible part $K(\omega)$ differs significantly from that in the $0<\mu<U$ range. As a consequence, DOS shapes are entirely different. The calculated results are shown in Fig.~\ref{Fig7}. The spectra demonstrate several well-resolved Landau subbands, even in the case of moderate and strong Hubbard repulsions. In the latter cases, we see the Mott gap separating low- and high-frequency Landau subbands, into which lower and upper Hubbard subbands divide. The number of Landau subbands forming lower and upper Hubbard subbands is equal to $\nu$ (in Fig.~\ref{Fig7}(c), the intensity suppression separating the second and third Landau subbands merges with the dip near the reabsorption frequency $\omega=-\mu$).
\begin{figure}[t]
\centerline{\resizebox{0.8\columnwidth}{!}{\includegraphics{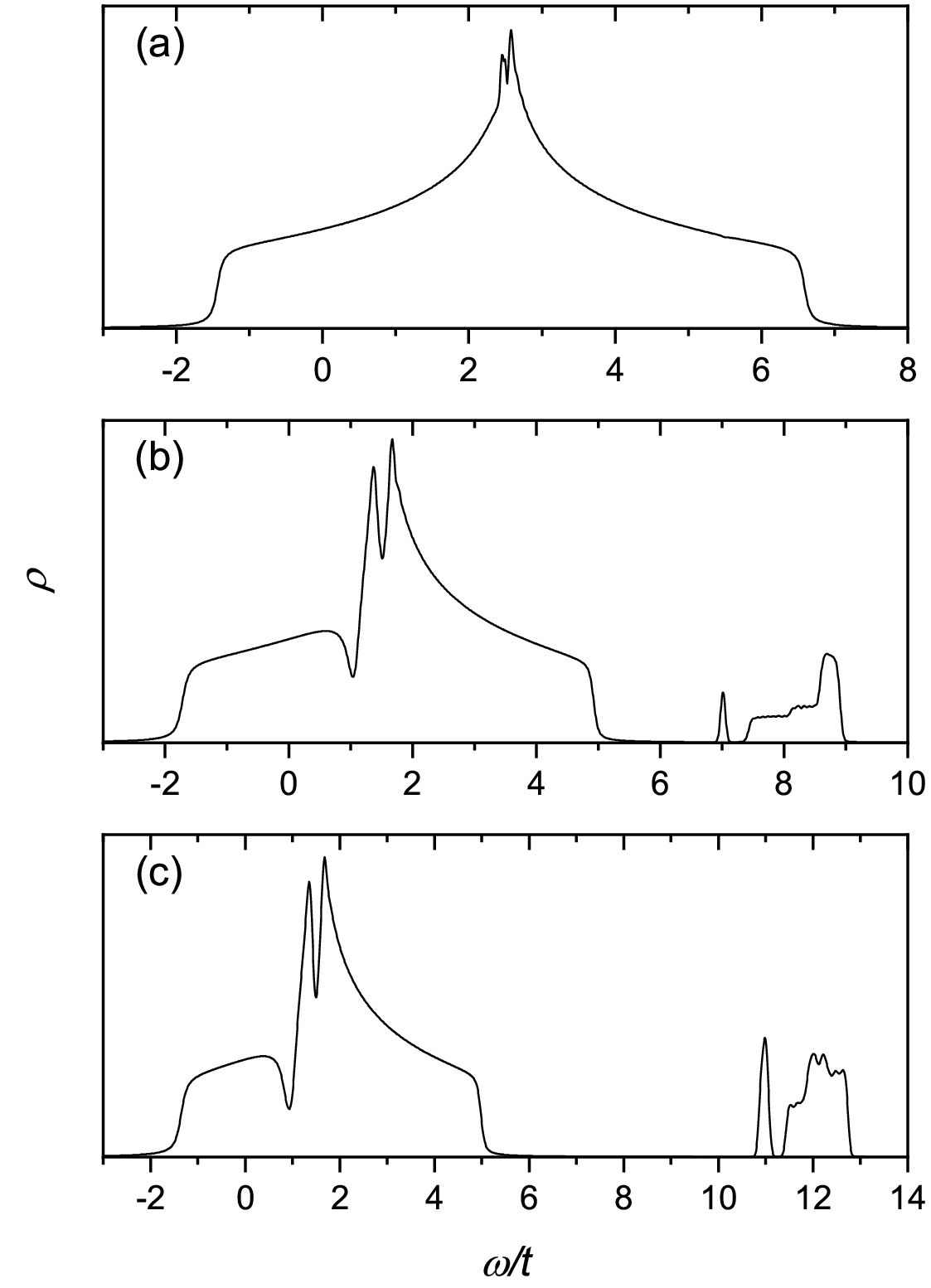}}}
\caption{Densities of states for $B=0$, $U=0.5t$, $\mu=-2.5t$ ($x=0.26$, a), $U=4t$, $\mu=-1.5$ ($x=0.35$, b), and $U=8t$, $\mu=-1.5t$ ($x=0.27$, c).} \label{Fig8}
\end{figure}
These results can be understood from the outcomes of Ref.~\cite{Sherman15b}. In that work, we considered the same problem using the simpler Hubbard-I approximation \cite{Hubbard}. Each Landau subband splits by the Hubbard repulsion into lower and upper Hubbard subbands in this approximation. Due to the narrowness of Landau subbands, the splitting occurs even for $U<U_c$. Consequently, the DOS contains the Mott gap separating an equal number of lower and upper Landau subbands. This picture is close to that shown in Fig.~\ref{Fig7}.

Notice that all DOSs in Fig.~\ref{Fig7} have dips at the reabsorption frequency $\omega=-\mu$. The intensity of peaks forming the upper Hubbard subband decreases with the diminution of $\mu$ from zero. The subband disappears as the Fermi level reaches the lower boundary of the spectrum, and for $U=4t$ and $8t$, the DOS shapes become similar to that shown in Fig.~\ref{Fig7}(a). This change of shapes is connected with the fact that the local interaction between electrons becomes increasingly rare with the decrease in their concentration.

For comparison, in Fig.~\ref{Fig8}, we show DOSs calculated for the same $U$ and $\mu$ without the magnetic field. For $U=4t$ and $8t$, the spectra demonstrate two Hubbard subbands. As for $B\neq0$, the intensity of the upper subband decreases with the diminution of $\mu$. The subband disappears as the Fermi level attains the lower edge of the spectrum, and the DOSs acquire a shape similar to that in Fig.~\ref{Fig8}(a). All DOSs have dips at frequencies $\omega=-\mu$.

\section{Conclusion}
In this work, we used the strong coupling diagram technique to investigate the zero-temperature density of states of the correlated electron system on a square lattice immersed in a perpendicular uniform magnetic field. In calculating the irreducible part, we considered two lowest-order diagrams from the SCDT series. On the one hand, in zero field, this approximation allows one to obtain the DOSs close to those calculated with an infinite series of SCDT ladder diagrams and in some other approaches. On the other hand, the approximation enables to perform the continuation to real frequencies analytically without using approximate numerical methods. Such continuation allows us to increase the resolution of obtained DOSs. The considered Hamiltonian has particle-hole symmetry. Therefore, we considered only electron concentrations equal to or less than unity. For $T=0$, the irreducible part and the related DOS change abruptly when the chemical potential crosses zero. This fact is connected with the sudden modification of the ground state of the Hubbard atom at this $\mu$ from empty to singly occupied state. For $\mu>0$, the Landau subbands in the DOS are well resolved at small Hubbard repulsions only. For moderate and large $U$, only slight DOS variations are retained from the merged subbands. All spectra contain gaps at the transfer frequencies of the Hubbard atom $\omega=-\mu$ and $U-\mu$. The gaps appear due to multiple electron reabsorptions on these frequencies. In contrast, for finite temperatures, the reabsorptions reveal themselves as spectral intensity suppressions at these frequencies. For $T=0$, widths of the reabsorption gaps grow as the chemical potential shifts from half-filling. At half-filling, as for $B=0$, the Mott gap opens at $U_c\approx6t$. Once this happens, small peaks appear below and above the gap. The peaks are retained at moderate doping. Their location depends on the magnetic field strength. This dependence leads to oscillations of the density of electron states on the Fermi level. Their frequency is low due to small areas of Fermi surface pockets arising with the Mott gap opening. This result can explain low quantum oscillation frequencies observed in lightly doped cuprates.

The DOSs calculated for $\mu<0$ are entirely different. For such chemical potentials, Landau subbands are well resolved not only for small repulsions, but for moderate and large $U$ as well. In later cases, Landau subbands form lower and upper Hubbard subbands with the Mott gap between them. This gap exists even for $U<U_c$. The numbers of the upper and lower Landau subbands are equal to $\nu$. The intensity of the upper subbands decreases as the Fermi level approaches the lower edge of the spectrum and finally disappears. This spectral change reflects that the local interaction between electrons becomes increasingly rare with decreasing concentration.


\begin{thebibliography}{99}
\bibitem{Shoenberg}D. Shoenberg, {\it Magnetic Oscillations in Metals} (Cambridge University Press, Cambridge, 1984)
\bibitem{Landau}L.D. Landau, E.M. Lifshitz, {\it Quantum Mechanics} (Pergamon Press, Oxford, 1965)
\bibitem{Brown}E. Brown, Phys. Rev. {\bf 133}, A1038 (1964). https://doi.org/10.1103/PhysRev.133.A1038
\bibitem{Langbein}D. Langbein, Phys. Rev. {\bf 180}, 633 (1969). https://doi.org/10.1103/PhysRev.180.633
\bibitem{Klitzing}K. v. Klitzing, G. Dorda, M. Pepper, Phys. Rev. Lett. {\bf 45}, 494 (1980). https://doi.org/10.1103/PhysRevLett.45.494
\bibitem{Thouless}D.J. Thouless, M. Kohmoto, M.P. Nightingale, M. den Nijs, Phys. Rev. Lett. {\bf 49}, 405 (1982). https://doi.org/10.1103/PhysRevLett.49.405
\bibitem{Tsui}D.C. Tsui, H.L. Stormer, A.C. Gossard, Phys. Rev. Lett. {\bf 48}, 1559 (1982). https://doi.org/10.1103/PhysRevLett.48.1559
\bibitem{Laughlin}R.B. Laughlin, Phys. Rev. Lett. {\bf 50}, 1395 (1983). https://doi.org/10.1103/PhysRevLett.50.1395
\bibitem{Sebastian}S.E. Sebastian, N. Harrison, G.G. Lonzarich, Rep. Progr. Phys. {\bf 75}, 102501 (2012). https://doi.org/10.1088/0034-4885/75/10/102501
\bibitem{Sherman15b}A. Sherman, Phys. Lett. A {\bf 379}, 1912 (2015). https://doi.org/10.1016/j.physleta.2015.05.023
\bibitem{Sherman22}A. Sherman, J. Low Temp. Phys. {\bf 209}, 96 (2022). https://doi.org/10.1007/s10909-022-02800-1
\bibitem{Markov}A.A. Markov, G. Rohringer, A.N. Rubtsov, Phys. Rev. B {\bf 100}, 115102 (2019). https://doi.org/10.1103/PhysRevB.100.115102
\bibitem{Vucicevic}J. Vu\v{c}i\v{c}evi\'c, R. \v{Z}itko, Phys. Rev. B {\bf 104}, 205101 (2021). https://doi.org/10.1103/PhysRevB.104.205101
\bibitem{Vladimir}M.I. Vladimir, V.A. Moskalenko, Theor.\ Math.\ Phys. {\bf 82}, 301 (1990). https://doi.org/10.1007/BF01029224
\bibitem{Metzner}W. Metzner, Phys.\ Rev. B {\bf 43}, 8549 (1991). https://doi.org/10.1103/PhysRevB.43.8549
\bibitem{Pairault}S. Pairault, D. S\'en\'echal, A.-M.S. Tremblay, Eur.\ Phys.\ J. B {\bf 16}, 85 (2000). https://doi.org/10.1007/s100510070253
\bibitem{Sherman18}A. Sherman, J. Phys.: Condens. Matter {\bf 30}, 195601 (2018). https://doi.org/10.1088/1361-648X/aaba0e
\bibitem{Grober}C. Gr\"ober C, R. Eder, W. Hanke, Phys.\ Rev. B {\bf 62}, 4336 (2000). https://doi.org/10.1103/PhysRevB.62.4336
\bibitem{Peierls}R. Peierls, Z. Phys. {\bf 80}, 763 (1933)
\bibitem{Wannier}G.H. Wannier, Rev. Mod. Phys. {\bf 34}, 645 (1962). https://doi.org/10.1103/RevModPhys.34.645
\bibitem{Sherman23}A. Sherman, Phys. Scr. {\bf 98}, 115947 (1923).  https://doi.org/10.1088/1402-4896/ad000b
\bibitem{Senechal02}D. S\'en\'echal, D. Perez, and D. Plouffe, Phys. Rev. B {\bf 66}, 075129 (2002). https://doi.org/10.1103/PhysRevB.66.075129
\bibitem{Slater}J.C. Slater, Phys. Rev. {\bf 82}, 538 (1951). https://doi.org/10.1103/PhysRev.82.538
\bibitem{Hubbard}J. Hubbard, Proc. Roy. Soc. (London) A281, 401 (1964). https://doi.org/10.1098/rspa.1964.0190
\end{thebibliography}
\end{document}